\RequirePackage{fix-cm}
\documentclass[smallextended]{svjour3}       % onecolumn (second format)
\usepackage{amsmath,amssymb}
\usepackage{xspace}

\smartqed  % flush right qed marks, e.g. at end of proof
\usepackage{graphicx}
\usepackage{mathptmx}      % use Times fonts if available on your TeX system

\newcommand{\ie}{{\it i.e.}\xspace}
\newcommand{\eg}{{\it e.g.}\xspace}

\newcommand{\nmax}{{n_{\text{max}}}\xspace}

\newcommand{\ave}[1]{\left\langle#1 \right\rangle}

\newcommand{\elabel}[1]{\label{eq:#1}}
\newcommand{\eref}[1]{(Eq.~\ref{eq:#1})}

\newcommand{\flabel}[1]{\label{fig:#1}}
\newcommand{\fref}[1]{Fig.~\ref{fig:#1}}

\journalname{}
\begin{document}

\title{Menger 1934 revisited 
%\thanks{I thank M. Gell-Mann for helpful comments and discussions. ZONlab Ltd. is acknowledged for support.}
}
%\subtitle{}


\author{Ole Peters}


\institute{O. Peters \at
              Department of Mathematics and Grantham Institute for Climate Change\\
Imperial College London\\
180 Queens Gate\\
SW7 2AZ\\
%              Tel.: +123-45-678910\\
%              Fax: +123-45-678910\\
              \email{ole@santafe.edu}           %  \\
}

\date{}

\maketitle

\begin{abstract}
Karl Menger's 1934 paper on the St. Petersburg paradox contains
mathematical errors that invalidate his conclusion that unbounded
utility functions, specifically Bernoulli's logarithmic utility, fail
to resolve modified versions of the St. Petersburg paradox.
\end{abstract}

\section{Preview and preliminaries}
In Section~\ref{The_St} the St. Petersburg paradox is
defined. Section~\ref{Ergodicity} motivates its recent resolution
using the concept of ergodicity, including Section~\ref{The_time}
where the ergodicity argument is related to a discussion in the
economics literature. Section~\ref{Bernoulli's} contrasts this with
D. Bernoulli's traditional 1738 resolution, and Section~\ref{Summary}
summarizes Menger's 1934 study, setting out its reception and
conceptual significance. Section~\ref{Error} focuses on Menger's
errors, with the central result in Section~\ref{Menger_overlooked}
that Menger overlooked a second divergence in the application of
Bernoulli's resolution to his modified paradox. Section~\ref{Menger's}
investigates Menger's game using the ergodicity resolution, and
Section~\ref{Summary_and} is a brief summary of the findings of the
present study.

\subsection{Text and notation} 
\label{Text}
Page numbers for \cite{Menger1934} refer to the English translation by
Schoellkopf and Mellon from 1967 \cite{Menger1967}. 

I will adopt the notation used there with the following
exceptions. Whereas Menger uses the same symbol $D$, and sometimes
$D_n$ for different functions of $n$, I use $D(n)$ for the general
case, $D_A(n)$ for the payout function used by Bernoulli, $D_B(n)$ for
that used by Menger, and $D_C(n)$ for a milder version of Menger's
game, see Table~\ref{games}. I introduce the symbol $P$ for the
ticket price, which -- significantly -- does not appear in Menger's
equations.

The printing of \cite{Menger1967} contains several typographical
errors, especially on p.~217, where the intended $e^{2^n}$ is
repeatedly misprinted as $e^{2n}$.  Also on this page,
$1/2^n$ is misprinted as $\frac{1}{2}/2^n$.  No further relevant
discrepancies with the original German text were found.

When this is convenient, monetary units are called dollars, \$, for
dimensional consistency, although the choice of unit or currency is
irrelevant. I will synonymously use the terms ``expectation value''
and ``ensemble average''.

Page numbers for \cite{Bernoulli1738} refer to the 1956 English
translation by Sommer.

\section{The St. Petersburg paradox}
\label{The_St}
N. Bernoulli, in a letter to Montmort in 1713 \cite{Montmort1713},
introduced lotteries of a certain type. D. Bernoulli, in a study
published by the Imperial Academy of Sciences in St. Petersburg,
considered the following specific case \cite{Bernoulli1738}: a fair
coin is tossed until the first heads event occurs. The number of coin
tosses necessary to arrive at this event is $n$, and the payout as a
function of $n$ is $\$D_A(n)=\$2^{n-1}$. We will refer to this as {\it
  game A}. The expectation value of the payout in {\it game A} is
given by the divergent sum
\begin{equation}
\ave{D_A(n)}=\sum_{n=1}^{\infty}\left(\frac{1}{2}\right)^n
2^{n-1}=\frac{1}{2}+\frac{1}{2}+\frac{1}{2}+ \cdots.  \elabel{payout}
\end{equation}
N. Bernoulli mentioned that he found this game ``curious''
\cite{Montmort1713}, p.~402.
%double-check page number in Montmort 1713 reprint XXX
Later researchers found it ``paradoxical'' that, in general,
individuals offered to purchase a ticket in this lottery are not
willing to pay very much for it, namely no more than a few dollars,
rather than their entire fortunes or indeed all the money they can
borrow. The surprise by these early researchers reflects the belief
that risky ventures may be judged by

\underline{\bf \it Criterion i:} \\ {\it a gamble is worth taking if
  the expectation value of the net change of wealth, here
  $\ave{D(n)}-P$, is positive.}

This criterion fails in the St. Petersburg paradox, in the sense that
there is no finite price $P$ at which it discourages
participation. {\it Criterion i} may be attributed to Huygens, who
wrote ``if any one should put 3 shillings in one hand without telling
me which, and 7 in the other, and give me choice of either of them; I
say, it is the same thing as if he should give me 5 shillings...''
\cite{Huygens1657}. Many resolutions were put forward, as reviewed \eg
in Menger(1934) and Samuelson(1977).

\section{Resolution based on Ergodicity}
\label{Ergodicity}
The present comment is written in view of the connection between
ergodic theory and the paradox. The St. Petersburg paradox rests on
the apparent contradicton between a positively infinite expectation
value of winnings in a game and real people's unwillingness to pay
much to be allowed to participate in the game. Bernoulli (1738),
p.~24, pointed out that because of this incongruence, the expectation
value of net winnings, {\it criterion i}, has to be ``discarded'' as a
descriptive or prescriptive behavioral rule. One can now decide what
to change about ``the expectation value of net winnings'': either
``the expectation value'' or the ``net winnings'' (or
both). Bernoulli, see Sec.~\ref{Bernoulli's}, chose to replace the net
winnings by introducing utility.
%and considered the expectation value of
%that. A disadvantage of this method is the introduction of the
%arbitrary utility function. 
An alternative resolution, motivated by the development of the field
of ergodic theory in the late 19th and throughout the 20th century,
replaces the expectation value (or ensemble average) with a time
average, without introducing utility. Details of this resolution can
be found in ~\cite{Peters2011}. Conceptually, ergodic theory deals
with the question whether expectation values, which can be thought of
as averages over non-interacting copies of a system (sometimes called
parallel universes), are identical to time averages, where the
dynamics of a single system are averaged along a time trajectory. It
is pointed out in \cite{Peters2011} that the system under
investigation, a mathematical representation of the dynamics of wealth
of an individual, is not ergodic, and that this manifests itself as a
difference between the ensemble average and the time average of the
growth rate of wealth. Historically, ergodic theory emphasizes the
relevance of the time average, which was discovered later.

%Bernoulli's resolution of 1738 predates the concept of ergodicity by
%more than one century \cite{Birkhoff1931,LebowitzPenrose1973}. The
%term ``ergodic'' was coined, according to \cite{Cohen1996}, p.~12, in
%\cite{Boltzmann1884}, and fully mathematized in 1931 by
%\cite{Birkhoff1931}. Menger, it seems, was not aware of this
%debate. For instance, he explains the meaning of expectation values
%using time-averages without pointing out that the two cannot generally
%be equated: ``[..]winnings and losses will be about equal, provided
%that the game is played a large number of times and that in each game
%the gambler risks his mathematical expectation [..]''
%\cite{Menger1934}, p.~211-212, footnote 2. This statement is correct,
%but it is meaningful only if the gambler is indeed able to risk his
%mathematical expectation in each game, \ie if the expectation value is
%small compared to his fortune. This is not the case in the
%St. Petersburg lottery, where the expectation value diverges, or -- in
%mathematical nomenclature -- does not exist and is of little practical
%relevance. As the amount of randomness (the size of fluctuations or
%strength of a noise term) diminishes, ensemble averages of dynamic
%properties (such as growth rates) often approach time averages, since
%in the case of no randomness all universes in the ensemble are
%identical and evolve deterministically.

To compute ensemble averages, only a probability distribution is
required, whereas time averages require a dynamic. This implies that
an additional assumption enters into the resolution in
\cite{Peters2011}. This assumption corresponds to the multiplicative
nature of wealth accumulation: any wealth gained can itself be
employed to generate further wealth, which leads to exponential
growth. The assumption is in general use, where the word ``general''
has the same meaning as in the following statement: in general, banks
and governments offer exponentially growing interest payments on
savings. This multiplicativity leads to a logarithm entering the
time-average growth rate. Different assumptions about the dynamics can
be envisioned, and these would lead to different functions appearing
in place of the logarithm. Consequences of replacing the logarithm
with generalized versions of it, in expressions very similar to
\eref{ergodic_criterion}, have been a topic of intense study in
recent years in statistical mechanics
\cite{HanelThurnerGell-Mann2011,Gell-MannTsallis2004}.

The treatment acknowledging the non-ergodicity of the system, assuming
multiplicative dynamics, produces

\underline{{\bf \it Criterion ii}}:\\ {\it a gamble is worth taking if
$\bar{g}$, the time-average exponential growth rate of a player's
wealth, is positive, where
\begin{equation}
\bar{g}=\sum_{n=1}^{\infty}\left(\frac{1}{2}\right)^n
\ln\left(\frac{W-P+D(n)}{W}\right).
\elabel{ergodic_criterion}
\end{equation}}
For a proof, see \cite{Peters2011}.

One method of computing both ensemble and time averages is to write
down for a finite number of sample members and a finite sampling time
an estimator of the quantity of interest. One then considers two
different limits: for the ensemble average the number of samples is
taken to infinity, and for the time average time is taken to
infinity. If the two limiting processes do not commute, then the
ensemble average is different from the time average, and the system is
manifestly not ergodic \cite{Peters2011,Peters2010}.

The encoding by the logarithm of effects of time in an ensemble
average (\eref{ergodic_criterion} is an ensemble average of the
logarithm) can be understood by writing the logarithm as a limit,
$\ln(x)=\lim_{q\to\infty}q(x^{1/q}-1)$. This limit corresponds, in the
present context, to the lottery being played repeatedly for an
infinitely long time in appropriately rescaled time units, see
\cite{Peters2011}. Thus, in \eref{ergodic_criterion}, which is an
ensemble average of a logarithmic growth rate (the time unit being one
lottery game), the time-limit is implicitly taken first. Since the two
limits do not commute in this case, this is not an ensemble average
but a time average.

%Expectation values can be relevant where essentially independent and
%initially identical systems exist and it is possible to combine, after
%some random event, the states of those systems. For instance, we could
%imagine an agreement among many players, offered tickets in
%independent St. Petersburg lotteries (with generally different
%outcomes), to share their winnings. But this is not the setup
%of N. Bernoulli's 1713 problem as described in his letter to Montmort
%\cite{Montmort1713}. 
There is no mention of repetitions of the game in N. Bernoulli's
letter, but one side of the paradox is human behavior, shaped by
evolution of individuals living through time, making decisions
repeatedly in risky situations. It is reasonable to assume that the
intuition behind the human behavior is a result of making repeated
decisions and considering repeated games.

%Whether or not one subscribes to the view that time averages are
%generally 
There is no doubt that in the St. Petersburg paradox the time average
is more relevant than the ensemble average: the ensemble average
diverges (does not exist). Nor does the interpretation of its
divergence as ``very large'' yield meaningful results. {\it Criterion
  i} must therefore indeed be discarded, as noted by Bernoulli. The
time average assuming the simplest dynamics, on the other hand,
resolves the paradox both mathematically and
behaviorally. Mathematically, there exists a finite price at which the
gamble should be rejected. Behaviorally, this finite price
approximately reflects people's choices, as is evident from
Bernoulli's resolution based on behavioral (albeit unsystematic)
observations, Sec.~\ref{Bernoulli's}, which for {\it game A} is
similar to {\it criterion ii}.

\subsection{The time argument in economics}
\label{The_time}
Following \cite{Kelly1956}, a debate regarding the use of time
averages began in the economics literature. Kelly had computed
time-average exponential growth rates in games of chance and argued
that utility was not necessary and ``too general to shed any light on
the specific problems'' he considered, \cite{Kelly1956}, p.~918. The
significance of time averages thus came very close to being fully
recognized in economics but the opportunity was missed, and the
connection to ergodic theory was not made. This is noteworthy because
from the perspective of the ergodicity debate the burden of proof is
on him who uses expectation values. Whereas the conceptual meaning of
a time average is clear -- resulting from a single system whose
properties are averaged over time, the use of an ensemble average has
to be justified, for instance by showing (for ergodic systems) that it
is identical to the time average. Alternatively, an ensemble can be a
good approximation if a large sample of essentially identical and
independent copies of the relevant system really exists, as was the
case in Boltzmann's studies \cite{Boltzmann1871b,Cohen1997}.

While the time argument was known
\cite{Kelly1956,Breiman1961,Markowitz1976}, its fundamental character
was not recognized. For instance, it was argued to be of limited
relevance because it assumes following a system's dynamics for an
infinite amount of time \cite{Samuelson1979}. Of course, time averages
are practically meaningful only if the real system has time to explore
the relevant part of its dynamical range. But the point of view of
ergodic theory emphasizes the opposite line of argument. To put it
provocatively, the ensemble average assumes an infinite number of
parallel universes. While real time scales are not infinite, they can
be large, whereas the real sample consists of exactly one system
(reality) and cannot be enlarged because that would imply the
absurdity of creating other universes, or -- more mildly -- copies of
the system.

Due to undetected errors and inaccuracies in \cite{Menger1934} it is
commonly stated in the economics literature that Menger proved {\it
  criterion ii} to be invalid, see below. Because it was not known
that {\it criterion ii} follows from considerations of ergodicity, and
that Menger's conclusions are at odds with a well-developed
mathematical field, his study was not subjected to sufficient
scrutiny, as will become apparent in Section~\ref{Error}.

%XXX move down.
%The difference between {\it criterion ii}, \eref{ergodic_criterion}
%and Bernoulli's {\it criterion iii}, \eref{Bernoulli_criterion} is not
%usually mentioned in the literature. Since Menger claimed to disprove
%Bernoulli, Menger thus contributed to the dismissal not only of
%Bernoulli's resolution, which does have an ``{\it ad hoc} character''
%\cite{Menger1934}, p.~217, but also to the dismissal of the ergodicity
%resolution, whose character is less {\it ad hoc}. It is for this
%reason that it is important to point out the flaws in Menger's
%argument.

\section{Bernoulli's 1738 resolution}
\label{Bernoulli's}
D. Bernoulli, writing about two centuries before the formulation of
the ergodic hypothesis, offered the following behavioral resolution of
the paradox: Since people care about their monetary wealth only
insofar as it is useful to them, he introduced the utility function
$U(W)$ that assigns to any wealth $\$W$ a usefulness. Specifically,
Bernoulli considered the function $U_B(W)=\ln(W)$. The quantity,
Bernoulli suggested, that people consider when deciding whether to
take part in the lottery is a combination of the expected gain in
their utility and the loss in utility they suffer when they purchase a
ticket. This leads to Bernoulli's

\underline{\bf \it  Criterion iii:} \\
{\it a gamble is worth taking if 
the following quantity is positive \cite{Bernoulli1738}, pp.~26--27:
\begin{equation}
\elabel{Bernoulli_criterion}
\ave{\Delta U_B^+} - \Delta U_{B}^{-}=\sum_{n=1}^{\infty}\left(\frac{1}{2}\right)^n
\ln\left(\frac{W+2^{n-1}}{W}\right) - \ln\left(\frac{W}{W-P}\right).
\end{equation}}
The first terms on either side of the equation represent the expected
gain in logarithmic utility, resulting from the payouts of the
lottery. This would be the net change in utility if tickets were given
away for free. The second terms represent the loss in logarithmic
utility suffered at the time of purchase, \ie after the ticket is
bought but before any payout from the lottery is received.  Notice
that Bernoulli did not calculate the expected net change in
logarithmic utility, which would be
\begin{equation}
\elabel{delta_u}
\ave{\Delta U_B}=\sum_{n=1}^{\infty}\left(\frac{1}{2}\right)^n
\ln\left(\frac{W+2^{n-1}-P}{W}\right).
\end{equation}
This expression, \eref{delta_u}, is mathematically identical to the
time-average growth rate, \eref{ergodic_criterion}. The expected net
change in logarithmic utility thus encodes in an ensemble average
information about time under multiplicative dynamics.

Bernoulli justifies his {\it criterion iii} as follows: ``in a fair
game the disutility to be suffered by losing must be equal to the
utility to be derived by winning'', \cite{Bernoulli1738}, p.~27. If
this leads us to believe that Bernoulli wanted to define a fair game
as one where the expected net change in utility is zero, then we will
conclude that he made an error in mathematizing this through
\eref{Bernoulli_criterion}. Bernoulli did not explicitly claim to
compute the expected net change in utility. Be it by mistake or
deliberately, he stated the belief that people tend to act according
to {\it criterion iii}, \eref{Bernoulli_criterion}. This criterion
states that, irrespective of future gains, one would never give away
one's entire fortune for a ticket in any game, as this would
correspond to an infinite loss in utility at the time of
purchase. This could be interpreted as a feeling of distrust. What if
the lottery is a scam? It can lead to nonsensical results if a future
gain greater than the player's wealth is certain, see
Section~\ref{The_worst}. The fact that we are forced to make
choices of interpretation is characteristic of the lack of clarity in
the debate from the outset. Any statement that is fundamentally
behavioral and seems incorrect to us can be interpreted either as a
mistake or as a reflection of a different belief regarding human
behavior.

The consensus in the literature on utility theory is that Bernoulli
meant to compute the expected net change in utility and made a slight
error. I have not been able to find an example -- other than
\cite{Menger1934} -- where the slight difference between Bernoulli's
expression \eref{Bernoulli_criterion} and the net change in
logarithmic utility, \eref{delta_u} was not implicitly
``corrected''. Already in \cite{Laplace1814}, p.~439--442, the
expected net change in utility is calculated, and the method is
ascribed to Bernoulli(1738), whereas \eref{Bernoulli_criterion} is not
mentioned. The book by \cite{Todhunter1865} follows Laplace, as do
modern textbooks, which state in one form or another that utility is
an object encoding human preferences in its expectation value, \eg
\cite{ChernoffMoses1959,Samuelson1983}.

\section{Summary and reception of Menger 1934}
\label{Summary}
Sections 1--5 and 7 of \cite{Menger1934} provide a review of
earlier treatments of the St. Petersburg paradox, Sections 2 and 10
clarifying comments regarding the nature of the paradox, and
Sections 8 and 9 a list of behavioral regularities with which
resolutions of the paradox should be consistent. In Section 5 Menger
criticized D. Bernoulli's 1738 resolution on the grounds of its ``{\it
  ad hoc} character'', referring to the arbitrariness of the chosen
utility function.

{\bf \underline{Menger's main finding:}}\\ {\it Menger's paper is best
  known for the conclusions in its Sections 5 and 6 that only bounded
  utility functions may be used if positively divergent expected net
  changes in utility, irrespective of ticket price, are to be avoided
  in games similar to the St. Petersburg lottery.}

This conclusion will be shown to be incorrect. Apart from calling
Bernoulli's solution ``{\it ad hoc}'' Menger avoids criticizing
Bernoulli conceptually and focuses his efforts on a mathematical
critique. As was shown in Section~\ref{Bernoulli's}, it is indeed
necessary to look carefully at Bernoulli's mathematical analysis, but
Menger, instead of discovering what was mathematically questionable in
\cite{Bernoulli1738}, was misled by precisely this questionable
detail to his incorrect conclusion in his Section 6 that only bounded
utility functions are permissible.

\subsection{Motivation for the present study}
The requirement of boundedness specifically rules out logarithmic
utility. It was shown in Section~\ref{Ergodicity}, however, that
logarithmic utility is mathematically equivalent to the conceptually
wholly different resolution based on ergodic theory. This point of
view provides a firm basis on which to erect a scientific formalism
that does not depend on psychological characteristics of human
beings. It allows, for instance, to define and compute optimal values
of leverage for investments \cite{Peters2010} and precipitates a new
notion of efficiency, which may be called ``stochastic market
efficiency'', based on dynamic stability arguments. This type of
efficiency has significant predictive power and has been corroborated
empirically \cite{PetersAdamou2011}. In this new context,
\cite{Menger1934} raises the question whether ergodic theory has to be
revised, Menger's argument is invalid or, of course,
\cite{Peters2010,Peters2011,PetersAdamou2011} are incorrect. Close
inspection shows that the slight inaccuracy in D. Bernoulli's 1738
computation of an expectation value propagated to the extreme context
of Menger's work, where it turns into a significant error and indeed
invalidates Menger's argument for the boundedness of utility
functions. Menger followed Bernoulli too closely but not carefully
enough, copying the central computation unquestioningly but
incompletely, whereas other researchers implictly corrected Bernoulli
\eg \cite{Laplace1814}, p.~440, \cite{Todhunter1865}, p.~221.

\subsection{Reception of Menger 1934}
Menger's detailed study \cite{Menger1934} is widely cited and
considered an important milestone in the development of utility theory
\cite{Markowitz1976,Samuelson1977,Arrow2009}.

It is interesting that the weakness of Menger's argument remained
undiscovered for at least 77 years (the argument was prepared in 1923,
presented in 1927 and published in 1934 \cite{Menger1934}). Menger's
paper was criticized for its polemicism by Samuelson: ``I myself would
be a bit more sparing in use of such phrases as `...he will not, if he
is sane...'  \cite{Menger1934}, p.~212; `...behavior of normal
individuals...'  \cite{Menger1934}, p.~213; `...no normal person would
risk his total fortune...'  \cite{Menger1934}, p.~217; `...behavior of
normal people...' \cite{Menger1934}, p.~222; `...it is clear that a
normal man will risk only a limited part of his total wealth to buy
chances in games' \cite{Menger1934}, p.~224'' \cite{Samuelson1977},
p.~48--49.  This writing style makes it difficult to detect the flaws
in Menger's argument. Linguistic clarity is lost where it is urgently
needed -- for instance, as we shall see below, the word ``risk'' needs
clarification, as do ``sane'' and ``normal''.

Credit is due to Arrow(1951) for writing clearly what we shall assume
to be what Menger had in mind (see
Section~\ref{Menger_overlooked}). Menger's implicit ruling-out of the
fundamental ergodicity solution was perceived as ``a drawback to the
empirical implementation of the expected-utility criterion, since many
of the most convenient forms (e.g., the logarithmic, the polynomials)
seem inadmissable because of their unboundedness'' \cite{Ryan1974},
p.~133, and Arrow commented that Menger's result ``clearly meets with
a good deal of resistance'', \cite{Arrow1974}, p.~136. Nonetheless,
also Arrow was misled and wrote recently, comparing to Bernoulli(1738)
``... a deeper understanding was achieved only with Karl Menger's
paper (1934)'', \cite{Arrow2009}, p.~93.

Despite his stylistic criticism Samuelson had enough confidence in
Menger's writing to conclude that ``After 1738 nothing earthshaking
was added to the findings of Daniel Bernoulli and his contemporaries
until the quantum jump in analysis provided by Karl Menger'',
\cite{Samuelson1977} p.~32, and further opined that ``Menger 1934 is a
modern classic that stands above all criticism'', p.~49. Several
renowned economists explicitly accepted Menger's argument, as
reflected in passages such as ``we would have to assume that $U$ was
bounded to avoid paradoxes such as those of Bernoulli and Menger''
\cite{Markowitz1976}, p.~1278, and to my knowledge no one explicitly
disagreed.

%Menger's 1934 study contains a mathematical error,
%possibly several errors, that will be pointed out here.
%We distinguish carefully between the correct
%and highly valuable parts of his study and those parts that must be
%regarded as flawed. 
%This is all the more important since the flaws do
%not contaminate much of the valid statements. Chiefly, they must be
%attributed to the state of utility theory in the 1930s and the
%arbitrariness of opinions regarding human behavior that is at the
%heart of this body of work.

\section{Errors in Menger's 1934 paper}
\label{Error}
Menger's error of conceptual type is the acceptance of an ensemble
average as a relevant criterion where only a single system
exists. With this acceptance he follows Bernoulli and inescapably
brings the discussion to a behavioral level. D. Bernoulli explicitly
states that the expectation value has to be ``discarded'' but he does
not do so. His 1738 paper does not use the expectation value of the
original gains in wealth of N. Bernoulli's work, but it still uses the
expectation value of the utility of payouts $\ave{\ln(D_A)}$. The
reader is referred to \cite{Peters2011} for further details of the
conceptual problem. In the following we will focus on Menger's
technical errors.

\subsection{The worst case in Menger's game}
\label{The_worst}
On p.~217 Menger(1934) introduced a modified St. Petersburg game, {\it
  game B}, whose payout for waiting time $n$ is given by the function
$D_B(n)=W\exp(2^n)-W$. Menger calls this a ``slightly modified game'',
presumably because, like the original {\it game A}, it offers large
payouts with small probabilities, and claims: ``.. it is obvious that,
even in the modified Petersburg Game, no normal person would risk his
total fortune or a substantial amount'', p.~217--218.

The tone of this statement motivates further probing. In this section
we clarify what Menger means by the word ``risk'' in this sentence. We
will thereby show that he probably overlooked the following
\vspace{.2cm}

{\bf \underline{Error 1:}}\\ {\it The worst case a player can suffer
  is a waiting time of $n=1$. But the payout in this worst case is
  $D_B(n=1)=W (\exp(2)-1) \approx 6.3 \times W$. In other words, in
  order to risk losing anything in the game, a player has to pay for a
  ticket more than 6.3 times his wealth, presumably by borrowing.}
\vspace{.2cm}

Did Menger mean this by ``risk'', \ie did Menger mean that ``no normal
person would'' borrow as much as or more than $\$W\times (\exp(2)-2)$
to buy a ticket, whereas a normal person may well borrow enough money
to pay three times his ``total fortune''? It seems that he did not
mean this, as we will see. It is most likely that Menger overlooked
the fact that the worst-case outcome in his game still poses no risk
to the person who pays $\$W$ for a ticket.

In the original {\it game A} there is little difference between
``risking'' and ``paying as a ticket price'' because the difference
between the worst-case net loss of $\$P-\$1$ (the amount at risk), and
the ticket price, $\$P$, is only \$1. In Menger's {\it game B}, on the
other hand, the difference between the worst-case net loss,
$\$P-\$W(\exp(2)-1)$, and the ticket price, $\$P$, is
$\$W(\exp(2)-1)$, \ie many times the individual's wealth.

In his Section 2, on p.~212, Menger wrote, referring to the original
{\it game A}: ``If B, who has an infinitely large expectation, would
be willing to pay any amount of money for the privilege of playing the
game, then this behavior would conform to his infinitely large
mathematical expectation [...] Not only will B not pay an infinitely
high price to play the game, since this is impossible, buy he will not
pay a very high price that he could afford. In any case, he will not,
if he is sane, risk all or even a considerable portion of his wealth
in a Petersburg game.''

This seems to indicate that if individuals were willing to pay the
highest price they can afford, Menger would have considered his paradox
resolved. Assuming that the highest price one can afford is one's
entire wealth ({\it i.e.} that borrowing is impossible), Menger's
statement that ``no normal person would risk his total fortune'' would
mean that by ``risk'' he meant ``pay as a ticket price'' and that he
thought no one would pay his total fortune for a ticket in his {\it
  game B}. But since the worst-case outcome is a gain greater than the
total initial fortune, there is no reason not to pay one's total
fortune. It seems that Menger did not notice by how much he
changed the original game.

This particular problem can be fixed by considering a different {\it
  game C}, that Menger did not propose, where the payout function is
$D_C(n)=\exp(2^n)$. But the more significant error, discussed in
Section~\ref{Menger_overlooked}, remains. The three different games
involved in the discussion are summarized in Table~\ref{games}.
\begin{table}
\caption{The three different games used here are defined by their
  payout functions $D(n)$, paid if in a sequence of fair coin-tosses
  heads first occurs on the $n^{\text{th}}$ toss.}
\label{games}       % Give a unique label
\begin{tabular}{lll}
\hline\noalign{\smallskip}
Introduced by & Payout $D(n)$ & Label  \\
\noalign{\smallskip}\hline\noalign{\smallskip}
N. Bernoulli 1713  & $2^{n-1}$ & {\it A} \\
K. Menger 1934 & $W \exp(2^n) -W$ & {\it B} \\
 & $\exp(2^n)$ & {\it C} \\
\noalign{\smallskip}\hline
\end{tabular}
\end{table}

\subsection{Menger overlooked a second divergence} 
\label{Menger_overlooked}
This section establishes the main finding, Menger's undiscovered
mathematical error mentioned in the introduction. In his attempt to
show that logarithmic utility fails to resolve the modified
St. Petersburg {\it game B}, Menger showed on p.~217 that the expected
payout in logarithmic utility is infinite,
\begin{equation}
\elabel{Menger_1}
\ave{\Delta U_B^+}=\sum_{n=1}^{\infty} \left(\frac{1}{2}\right)^n \ln\left(\frac{W+W\exp(2^n)-W}{W}\right)=\sum_{n=1}^\infty 1.
\end{equation}
Menger continues with the statement: ``it is obvious that, even in the
modified Petersburg Game, no normal person would risk his total fortune
or a substantial amount'' \cite{Menger1934}, p.~217--218.

It was pointed out in Section~\ref{The_worst} that depending on the
meaning of the word ``risk'' this may be incorrect. The problem to be
discussed now is that Menger does not actually use Bernoulli's {\it
  criterion iii}, \eref{Bernoulli_criterion}. Nor does he compute the
expected net gain in logarithmic utility, which equals the time
average growth rate under the dynamics discussed above, {\it criterion
  ii} \eref{ergodic_criterion}. He only considers the first term of
{\it criterion iii}, \eref{Bernoulli_criterion}.

It seems that Menger's reasoning went as follows: The expected gain in
utility at zero ticket price is infinite, wherefore any loss in
utility resulting from a finite ticket price must be negligible. This
reveals a linear way of thinking -- it would be correct if utility
were linear (in which case the problems of {\it criterion i} would
resurface). Menger's analysis amounts to the introduction of 

\underline{\it \bf Criterion iv:}\\ {\it a gamble is worth taking,
  irrespective of the ticket price, if the expected utility-payout,
  \eref{Menger_1}, is positive. }

For instance, any real-life lottery, where a ticket is purchased and
with a small probability something is won, is recommended by this
criterion. See Table~\ref{criteria} for all criteria.

\begin{table}
\caption{The straw-man {\it criterion i} considers the expected net
  change in wealth. {\it criterion ii} considers time-average growth
  rates which are mathematically identical to expected net changes in
  utility. It resolves games {\it A}, {\it B} and {\it C}: there is a
  finite ticket price at which it discourages participation.  While
  the utility interpretation of {\it criterion ii} is purely
  behavioral and can only be judged by experiment, the time
  interpretation allows predictions and judgments of when this
  resolution does not apply in reality. The assumption of games played
  at an arbitrarily high frequency becomes crucial here and is not
  satistfied in real life. {\it Criterion iii} is precisely
  D. Bernoulli's resolution and implies the recommendation never to
  pay one's entire wealth for a ticket. Menger's {\it criterion iv}
  shows that he only computed the first term of Bernoulli's {\it
    criterion iii}. Since {\it criterion iv} does not include the
  ticket price, it fails for all three games.}
\label{criteria}
\begin{tabular}{lcll}
\hline\noalign{\smallskip}
Introduced by & Criterion & Games resolved  & Label\\
\noalign{\smallskip}\hline\noalign{\smallskip}
Huygens 1657 & $\underbrace{\ave{D(n)}-P}_{\text{expected net wealth change}}$& none & {\it i}\\
&&&\\
Laplace 1814 & $\underbrace{\ave{\ln(W+D(n)-P)}-\ln(W)}_{\text{expected net utility change = time ave. growth rate}}$ & {\it A, B, C} & {\it ii}\\
&&&\\
D. Bernoulli 1738  & $\underbrace{\ave{\ln(W+D(n))}-\ln(W)}_{\text{expected utility gain}}-\underbrace{[\ln(W)-\ln(W-P)]}_{\text{utility loss at purchase}}$ & {\it A, B, C} & {\it iii}\\
&&&\\
Menger 1934 & $\underbrace{\ave{\ln(W+D(n))}-\ln(W)}_{\text{expected utility gain}}$ & none & {\it iv}\\
\noalign{\smallskip}\hline
\end{tabular}
\end{table}

For clarity we turn to \cite{Arrow1951}, who summarized Menger as follows:
``Let $U(x)$ be the utility derived from a given amount of money $x$,
and suppose that $U(x)$ increases indefinitely as $x$ increases
[e.g. if $U(x)=\log(x)$, as in Bernoulli's own theory]. Then, for
every integer $n$, there is an amount of money $x_n$, such that
$U(x_n)=2^n$. Consider the modified St. Petersburg game in which the
payment when a head occurs for the first time on the $n$th toss is
$x_n$. Then, clearly, the expected utility will be the same as the
expected money payment in the original formulation and is therefore
infinite. Hence, the player would be willing to pay any finite amount
of money for the privilege of playing the game. This is rejected as
contrary to intuition.''

Samuelson agrees: ``...Menger shows one can always fabricate a
Super-Petersburg game [...] to make the compensation Paul needs again
be infinite.'' \cite{Samuelson1977}, p.~32.

Both Arrow's and Samuelson's phrasings of Menger's central result
agree with my reading of Menger's statements. These statements are
incorrect because of
\vspace{.2cm}

{\bf \underline{Error 2:}}\\ 
{\it Menger overlooked that the logarithm $\ln(W-P)$ diverges
  negatively at a finite value of $P$.}  
\vspace{.2cm}

If he did not overlook this divergence, then he overlooked Bernoulli's
second term altogether. 
%At this point Menger does something he
%explicitly wanted to avoid: he considers the divergent \eref{Menger_1}
%paradoxical, whereas he writes on p.~212, referring to the divergence
%in {\it game A}: ``What usually is considered paradoxical or at least
%requiring explanation is the fact that the mathematical expectation
%value of the person playing the game is not a finite sum, but
%infinite. In the present paper, however, the position will be taken
%that this infinitude in itself is not paradoxical.''
While it is correct that the first term of \eref{Bernoulli_criterion}
diverges positively, making the game appear attractive, the second
term diverges negatively as $P\to W$. Samuelson's statement,
paraphrasing Menger, is therefore incorrect, and Arrow's is correct
only up to ``Hence, the player would be willing...''

Menger's game produces a case of competing infinities. For values
$P\geq W$, Bernoulli's {\it criterion iii} is not defined. We have to
compare the divergences that lead to the infinities to understand what
this undefined region signifies. To this end we consider a finite
lottery, identical to {\it game B} except that if heads does
not occur in $\nmax$ coin tosses, the game ends, and the lottery
returns the ticket price\footnote{Laplace introduced an $\nmax$ in
  {\it game A}, but chose to end the lottery and return nothing after
  $\nmax$ successive tails events\cite{Laplace1814}, p.~440.}. For
any finite value $\nmax$, the corresponding partial sum can
be compensated for by a ticket price close enough to the wealth of a
person, so that {\it criterion iii} does not favor participation in the
lottery
\begin{equation}
\elabel{partial_sum}
\forall n_{\text{max}}<\infty \hspace{.2cm} \exists \hspace{.2cm} P_0<W : \sum_{n=1}^{n_{\text{max}}}\left(\frac{1}{2}\right)^n
2^n - \ln\left(\frac{W}{W-P_0}\right)=0.
\end{equation}
To ensure that the criterion remains non-negative (recommending
participation) up to exactly $P_0=W$, events of zero probability must
be taken into account. It is in this sense that the region where
\eref{Bernoulli_criterion} is undefined corresponds to the
recommendation not to buy a ticket, and the diverging expectation
value of the utility change resulting from the payout is dominated by
the negatively diverging utility change from the purchase of the
ticket.

The dominance of the divergence of the logarithm over the diverging
sum can be expressed mathematically in many different ways. For
instance, \eref{partial_sum} can be re-written as
\begin{equation}
\elabel{nmax}
\nmax(P_0)=\ln\left(\frac{W}{W-P_0}\right), 
\end{equation}
implying that the maximum allowed number of coin tosses, $\nmax$, has
to approach infinity in order to counter-balance a price close to the
finite value $W$, see \fref{P_of_n}.  Inverting \eref{nmax} yields
$P_0(\nmax)=W\left(1-\exp\left(-\nmax\right)\right)$: no matter how
many terms, or coin-tosses, are taken into account, the ticket price
never has to be greater than $W$ to compensate for the diverging sum.

\begin{figure*}
  \includegraphics[width=0.75\textwidth]{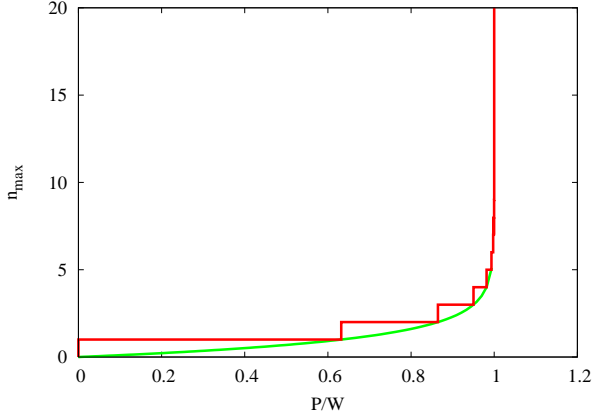}
\caption{Mininum (integer) number of coin tosses, $\nmax$, that need
  to be allowed in {\it game B} to render D. Bernoulli's {\it
    criterion iii} positive for a given price (red line), and the real
  number (green line) that renders the criterion exactly zero. The
  limiting value is $P_0=W$, where an infinity of tosses is
  needed. Menger's {\it game B} is very profitable -- even if
  $P_0=W/2$ and the number of coin tosses is limited to one, the
  cautious {\it criterion iii} recommends playing.}
\flabel{P_of_n} % Give a unique label
\end{figure*}

Far from suggesting recklessly risky behavior as {\it criterion i}
does, Bernoulli's {\it criterion iii} can be criticized for being too
cautious. As we have seen, the worst-case outcome of Menger's game
(with $\nmax=\infty$) is a payout of some $630\%$ of the player's
initial wealth. At a ticket price of twice the player's wealth, the
worst-case net result is still an increase in wealth of $430\%$, but
Bernoulli's {\it criterion iii} discourages participating.

\section{Menger's game and ergodicity}
\label{Menger's}
In {\it game B} the difference between {\it criterion ii} and {\it
  criterion iii} becomes visible. While Bernoulli's {\it criterion iii}
recommends buying a ticket as long as it is less expensive than one's
total wealth ($P<W$), {\it criterion ii} suggests to buy a ticket as
long as it is not possible to lose one's entire wealth ($P<W
\exp(2)$).

{\it Criterion ii} is in this sense much riskier than Bernoulli's,
which raises the issue of the relationship between mathematics and
reality. Is it really advisable to pay for a ticket $\$W
\exp(2)-\$\epsilon$, where $\epsilon$ is an arbitrarily small number?
With 50\% probability heads occurs on the first toss, and the player
ends up penniless (except for $\$\epsilon$). Mathematical statements
are correct in a sense that statements of other type are not. The
price we pay for the certainty inherent in mathematical statements is
that they refer to objects that don't exist in an everyday sense of
the word ``exist''. It would be na\"ive to use any mathematical
criterion as a guide for decisions without questioning the criterion's
conceptual meaning and the axioms upon which its derivation
rests. Applicability, or relevance to the real world, pertains only
insofar as the axioms and their consequences are a reflection of the
real world. In the case of the expectation-value {\it criterion i},
this ceases to be the case when fluctuations become significant and
effects of time, of events occuring in sequence, can no longer be
ignored. The ergodicity resolution, {\it criterion ii}, assumes that
equivalent lotteries can be played arbitrarily often, and this breaks
down in Menger's game -- games of his type are atypical in the real
world. If we come across such a game, it is a once-in-a-lifetime
opportunity. Mathematics cannot offer much help in this situation. It
makes sense of random events only by embedding them in an ensemble of
similar events, which may reside in parallel systems or within
time. In the single-event setup, mathematics must shrug its shoulders
and admit defeat in the face of a moral decision.

There is nothing wrong with the mathematics in the ergodicity
resolution \cite{Peters2011}. If we are really offered Menger-type
games to be played at an arbitrarily high frequency, so that we can
play arbitrarily often in our finite lives, then {\it criterion ii}
will be practically meaningful.
%No mathematical criterion, or quantitative description of any kind,
%will always be applicable. A mature use of mathematics requires
%specifying limits of applicability. The sign of the time-average
%growth rate will cease to be a sensible criterion in extreme cases. 
It is important to note that the time average growth rate (or expected
net change in logarithmic utility) in {\it game B} does not fail to
resolve Menger's paradox mathematically (unlike the expectation value
in the original {\it game A}). There is a finite price, $P=W\exp(2)$,
from which {\it criterion ii} discourages playing {\it game B}.

With payouts $D_C(n)=\exp(2^n)$ ({\it game C}) Bernoulli's {\it
  criterion iii} recommends to abstain from the gamble for ticket
prices $P\geq W$.  {\it criterion ii} recommends to abstain if paying
the ticket price poses the risk of bankruptcy, \ie if $P\geq
W+\exp(2)$.  {\it Criteria i} and {\it iv} fail for {\it game C} in
the sense that {\it criterion i} recommends paying any finite price,
though not necessarily an infinite price, and {\it criterion iv}
recommends paying even an infinite price.

\section{Summary}
\label{Summary_and}
In 1738 D. Bernoulli made an error, or approximation -- insignificant
in his context -- when he computed the expected net change in
logarithmic utility. Perceiving this as an error, Laplace in 1814 and
later researchers corrected it implicitly without mention. In
1934 Menger unwittingly re-introduced Bernoulli's error and introduced
a new error by neglecting a diverging term. Throughout the twentieth
century, Menger's incorrect conclusions were accepted by prominent
economists such as Arrow, Samuelson and Markowitz, although at least
Arrow and Ryan noticed and struggled with detrimental consequences of
the (undetected) error for the developing foramlism. The equivalence
of logarithmic utility and time-averaging in the non-ergodic system
allows a physical interpretation. This provides a basis for the
intuition that led to the discovery of Menger's error, as discussed
above.

The discussion surrounding Menger's 1934 paper deals with 3 different
games, defined by their payouts $D(n)$, Table~\ref{games}; and four
different criteria to evaluate them, Table~\ref{criteria}.

%\begin{table}
%\begin{tabular}{lll}
%\hline\noalign{\smallskip}
%\end{tabular}
%\end{table}

%criterion i
{\bf Criterion i}\\
As noted by N. Bernoulli, use of the expectation-value {\it criterion
  i} produces nonsensical recommendations in {\it game A}, which is
true also for {\it games B} and {\it C}. It is important to remember
that there is no {\it a priori} reason for the expectation value to be
a meaningful quantity because it is conceptually based on an ensemble
of systems, whereas the paradox deals with only one system.
%Taking a
%time average with the simplest dynamics, as required by the
%non-ergodicity of the system, gives sensible recommendations in the
%case of the original St. Petersburg paradox, {\it game A}. 
The failure of {\it criterion i} is both mathematical (there is no
finite price that should not be paid for a ticket) and conceptual
(real people do not behave this way).

%criterion ii
{\bf Criterion ii}\\
{\it Criterion ii} expresses the irreversibility of time and the
non-ergodicity of the system, and it can be phrased mathematically
identically to the requirement of an expected net increase in
logarithmic utility. It resolves {\it game A} similarly to {\it
  criterion iii}. {\it Games B} and {\it C} are mathematically
resolved: no price $P\geq D(n=1)+W$ should be paid for a
ticket. Behaviorally, the situation is more complicated. The
assumption of infinite repeatability becomes important, and this is
not usually realistic. Real people would thus be ill-advised to take
this criterion literally in {\it games B} and {\it C}.
%For {\it game A} the criterion is in line with real
%people's behavior. 
Since the criterion is rooted in ergodic theory, or in the physical
argument of time-irreversibility, it does not suffer from the
arbitrariness of {\it criterion iii}.

%criterion iii
{\bf Criterion iii}\\
Menger claimed that Bernoulli's original {\it criterion iii} fails in
the same sense in the modified {\it game B}. But that is
incorrect. {\it Criterion iii} fails neither mathematically (no
$P\geq W$ should be paid for a ticket) nor behaviorally (real
people may well be limited by a no-borrowing constraint, or unwilling
to pay more than their wealths). The paradox with {\it game C} is
analogously resolved. {\it Criterion iii} may be criticized for being
too strict -- there is no fundamental reason not to pay more than
$\$W$ if the minimum payout $\$D(n=1)>\$W$. It may also be
criticized for being somewhat arbitrary -- why is a guaranteed return
of 540\% acceptable but not one of 520\%?

%criterion iv
{\bf Criterion iv}\\
Menger's {\it criterion iv} fails to resolve all games, even the
original {\it game A}, because it does not include the ticket price,
which appears in the second term of {\it criterion iii},
\eref{Bernoulli_criterion}. Irrespective of the ticket price, any
gamble with a positive expected payout (not net-payout) should be
taken, and no gamble with a negative expected payout should be taken
(even if one were paid for participating). Menger was certainly under
the impression that he was using Bernoulli's {\it criterion iii},
although it is unclear whether he was aware that this is different
from {\it criterion ii}. By using {\it criterion iv} Menger
misrepresents Bernoulli, and his conclusions regarding both {\it
  criteria ii} and {\it iii} are mathematically incorrect. His
rejection of unbounded utility functions is mathematically unfounded,
based on errors. This is important because different utility functions
can be shown to correspond to different types of dynamics. The
logarithm (an unbounded function) corresponds to exponential growth,
one of the simplest dynamics and a good model for many natural
processes.

%\section{Section title}
%\label{sec:1}
%Text with citations \cite{RefB} and \cite{RefJ}.
%\subsection{Subsection title}
%\label{sec:2}
%as required. Don't forget to give each section
%and subsection a unique label (see Sect.~\ref{sec:1}).
%\paragraph{Paragraph headings} Use paragraph headings as needed.
%\begin{equation}
%\end{equation}

%\begin{figure}
%  \includegraphics{example.eps}
%\end{figure}
%

\begin{acknowledgements}
I thank M. Gell-Mann for helpful comments and discussions. ZONlab
Ltd. is acknowledged for support.
\end{acknowledgements}

\bibliographystyle{spbasic}      % basic style, author-year citations
\bibliography{/Users/obp48/Downloads/bibliography}

%\begin{thebibliography}{}
% etc
%\end{thebibliography}

\end{document}